\documentclass[twocolumn,showpacs,preprintnumbers,amsmath,amssymb]{revtex4}

\usepackage{latexsym}%
\usepackage{graphicx}
\usepackage{dcolumn}
\usepackage{bm}

\begin{document}

\voffset= 1.0 truecm 

%
\newcommand{\bea}{\begin{eqnarray}}
\newcommand{\eea}{\end{eqnarray}}
\newcommand{\be}{\begin{equation}}
\newcommand{\ee}{\end{equation}}
%
\newcommand{\xbf}[1]{\mbox{\boldmath $ #1 $}}

\title{Pion couplings of the $\Delta(1232)$}

\author{Alfons J. Buchmann}
\affiliation{Institute for Theoretical Physics,
University of T\"ubingen, D-72076 T\"ubingen, Germany}
\email{alfons.buchmann@uni-tuebingen.de}
\author{Steven A. Moszkowski} 
\affiliation{Department of Physics and Astronomy, University of California, 
Los Angeles, CA 90095-1547, USA}
\email{stevemos@ucla.edu}

\pacs{11.30.Ly, 12.38.Lg, 13.75Gx, 14.20.Gk}

\begin {abstract}
We calculate the strong couplings of pions to the $\Delta(1232)$ resonance 
using a QCD parameterization method that includes in addition to the usual 
one-quark also two-quark and previously uncalculated three-quark operators.
We find that three-quark operators are necessary to obtain 
results consistent with the data and other QCD based baryon structure models. 
Our results are also in quantitative agreement 
with a model employing large $D$ state admixtures to the $N$ and $\Delta$ 
wave functions indicating that the $\pi N$ and $\pi \Delta$ couplings are 
sensitive to the spatial shape of these baryons.

\end{abstract}

\maketitle

\section{Introduction}

Since its discovery~\cite{fer52} in pion$\,(\pi)$-nucleon$\,(N)$ scattering
the lowest excited state of the nucleon with spin $S=3/2$ and isospin 
$T=3/2$, called $\Delta(1232)$, has been important 
both for an understanding of nucleon ground state structure~\cite{pas07} 
and the nucleon-nucleon interaction~\cite{mac89}. 
In sufficiently energetic $N N$ collisions one or both nucleons may be 
excited to the $\Delta(1232)$, a process that makes a significant 
contribution to  many nuclear phenomena~\cite{mos87}, for example,
electromagnetic properties of light nuclei~\cite{are74}, three-nucleon
forces~\cite{epe08}, and the binding energy of nuclear matter~\cite{fer90}.
In addition, $\Delta$ degrees of freedom are needed to explain the
empirical cross sections for the $\pi N \to \pi N$~\cite{fet01}, 
$\pi N \to \pi \pi N$~\cite{arn79,Kam04}, and  
$NN \to NN \pi \pi$~\cite{sko11} reactions including a novel
resonance structure in the $p\, n \to d \pi^0 \pi^0$ 
channel~\cite{adl11}.

\begin{figure}
\resizebox{0.48\textwidth}{!}{
\includegraphics{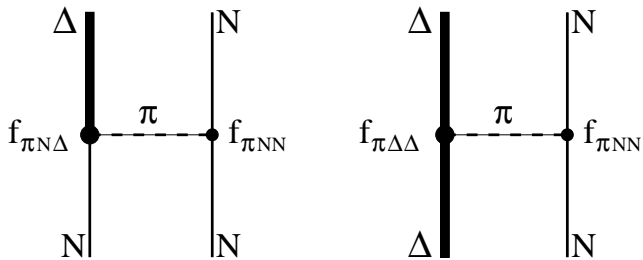}}
\caption{\label{fig:deltacoup}
Strong coupling of the pion to the nucleon ($N$) and $\Delta$-isobar 
($\Delta$). The $\pi NN$, $\pi N \Delta$, and $\pi \Delta \Delta$ 
coupling constants are denoted as $f_{\pi NN}$, $f_{\pi N \Delta}$, 
and $f_{\pi \Delta \Delta}$. The corresponding interaction vertices are 
represented as black dots.}
\end{figure}

However, a quantitative assessment of the role of $\Delta$ degrees of 
freedom in nuclear physics has remained difficult 
due to the lack of detailed knowledge 
of even basic $\Delta$ properties. For example, 
the widely used additive quark model, which is based on the assumption 
that observables can be calculated using a sum of one-quark operators,
underpredicts the experimental $N \to \Delta$ 
transition magnetic moment by about $30\%$~\cite{dal66} but 
slightly overpredicts the $\Delta^+$ magnetic moment~\cite{kot02}.
Another example involves the strong coupling constants 
$f_{\pi N \Delta}$ and $f_{\pi \Delta \Delta}$. 
The former determines the decay rate $\Delta \to N + \pi$, 
while the latter fixes the $N \Delta$ interaction strength 
in nuclei (see Fig~\ref{fig:deltacoup}). 
Here again, the additive quark model underpredicts the empirical
$N \to \Delta$ transition coupling
$f_{\pi N \Delta}$ by $20\%$, whereas it appears to 
overpredict the double $\Delta$ coupling $f_{\pi \Delta \Delta}$. 
A resolution of these discrepancies is necessary for a quantitative 
description of $\Delta$ degrees of freedom in nuclei~\cite{gre76,are78,eri88}.
   
Previously, the strong $\pi \Delta$ couplings were calculated 
with a QCD parameterization method, in which in addition to  
one-quark operators (additive quark model), 
two-quark operators were taken into account~\cite{hen00}.   
It was shown that two-quark contributions amount to 
a 20$\%$ increase of $f_{\pi N \Delta}$ with respect to the 
additive quark model results.
Furthermore, the following relation between the $\pi N$, $\pi N \Delta$ 
and $\pi \Delta \Delta$ 
couplings was derived
\begin{equation}
\label{rel1}
f_{\pi^0 p p}- \frac{1}{4} \, f_{\pi^0 \Delta^+\Delta^+} =
\frac{\sqrt{2}}{3} \, f_{\pi^0 p \Delta^+}.   
\end{equation}
This relation connects the elusive
$f_{\pi^0 \Delta^+\Delta^+}$ to the better known
 $f_{\pi^0 p\Delta^+}$ and $f_{\pi^0 p p}$ couplings.
Throughout this paper we use the normalization 
conventions of Ref.~\cite{Bro75} for the one-quark contributions, i.e., 
$f_{\pi \Delta\Delta}= (4/5)\, f_{\pi NN}$ and
$f_{\pi N \Delta}= (6\sqrt{2}/5)\, f_{\pi NN}$~\cite{comment1}. 
These theoretical relations, which are based on the additive quark model 
(see sect. 3), satisfy Eq.(\ref{rel1}).
On the other hand, if the empirical relation 
$f_{\pi^0 p \Delta^+}=2.1 \, f_{\pi^0 p p}$  
is used in Eq.(\ref{rel1}), one obtains 
$f_{\pi^0 \Delta^+\Delta^+}=0.04 \,f_{\pi^0 pp}$. 
However, from the viewpoint of QCD sum rules~\cite{zhu00} and 
$1/N_c$ expansion~\cite{das94}  
both $f_{\pi^0 \Delta^+\Delta^+}$ 
and $f_{\pi^0 p p}$ should be of the same order of 
magnitude.

The main purpose of this paper is to investigate if 
the inclusion of three-quark operators can resolve the discrepancies
between theory and experiment for the $\pi \Delta$ couplings.
Another motivation for this study comes from the work of 
Abbas~\cite{aba91}, who found that large $D$-wave admixtures 
in the $N(939)$ and $\Delta(1232)$ quark wave functions 
reduce the additive quark model result for $f_{\pi\Delta\Delta}$ 
by $20\%$ while they increase $f_{\pi N\Delta}$ by about the same percentage. 
This indicates that the 
strong $\pi \Delta$  couplings are sensitive to the spatial shape 
of the quark distribution in the nucleon and its first excited state. 
Therefore, it is of interest to study whether the two- and three-quark 
terms have an analogous effect on the strong $\pi \Delta$ couplings 
and may thus be interpreted as describing degrees of freedom 
leading to nonspherical geometrical shapes of the $N$ and $\Delta$ baryons. 

\section{Method}
As in our previous work we use a general parametrization (GP)
method developed by Morpurgo and described
in more detail in Refs.~\cite{Mor89,Mor92,Dil99} to calculate the strong 
pion couplings.
The most general expression for the corresponding operator
${\cal O}$ that is compatible with the space-time and inner 
QCD symmetries is a sum of one-, two-, and three-quark operators 
in spin-flavor space multiplied by {\it a priori} unknown constants
(called $A_1$, $A_2$, and $A_3$ below), 
which parametrize the orbital and color space matrix elements.
Empirically, a hierarchy in the importance
of one-, two-, and three-quark operators is found.
This fact can be understood
in the $1/N_c$ expansion where
two- and three-quark operators 
are usually suppressed by powers of $1/N_c$ and $1/N_c^2$
respectively compared to one-quark operators~\cite{Ncpapers}.
The two- and three-quark contributions are an effective description of 
gluons and quark-antiquark degrees of freedom that have been eliminated
from the QCD wave function~\cite{Mor89}.

For the strong $\pi N$ and $\pi \Delta$ couplings one-, two-, and three-quark 
axial vector operators are defined as
\bea
\label{operators}
{\mathcal{O}}_1 & = & A_1\, \sum_i \tau_3^i  \sigma_z^i, \nonumber \\
{\mathcal{O}}_2 & = & A_2\, \sum_{i\neq j} \tau_3^i \sigma_z^j  \nonumber \\
{\mathcal{O}_3} & = & A_3\, \sum_{i\neq j \neq k} \, \tau_3^i\, \sigma_z^i\, 
{\xbf{\sigma}}^j \cdot {\xbf{\sigma}}^k,
\eea
and the total operator reads
\begin{equation}
\label{quarkop}
{\mathcal{O}}={\mathcal{O}}_1 + {\mathcal{O}}_2 + {\mathcal{O}}_3.
\end{equation}
Here, ${\xbf{\sigma}}^i$ and ${\xbf{\tau}}^i$ are the spin and isospin
operators of quark $i$.

In Ref.~\cite{hen00} we briefly discussed why the 
two-body operator ${\cal O}_2$
in Eq.(\ref{operators}) is unique. 
With respect to the three-quark operator ${\cal O}_3$ 
the reader may wonder why other three-quark operators, for example
\bea
\label{other_operators}
\widetilde{\mathcal O}_3 & = &  \sum_{i\neq j \neq k} \, \tau_3^i\, 
\left \lbrack 
{\xbf{\sigma^i}} \times {\xbf{\sigma}}^j \times {\xbf{\sigma}}^k 
\right \rbrack_z \\
\widehat{\mathcal O}_3 & = &  \sum_{i\neq j \neq k} \, \tau_3^i\, 
\sigma^j_z \, \, {\xbf{\sigma}}^i \cdot {\xbf{\sigma}}^k \nonumber
\eea
can be excluded from the list of permissible operators.
It turns out that the operator $\widetilde{\cal O}_3$ 
is identical to zero when summed over quark indices. Furthermore, the operator
$\widehat{\cal O}_3$ has for the spin-flavor symmetric $N$ and $\Delta$ states 
considered here, the same matrix elements as the two-body 
operator ${\cal O}_2$ in Eq.(\ref{operators}) so that its effect 
is already included~\cite{comment3}. This is an example of an SU(6) 
operator reduction rule. More generally, SU(6) operator reduction 
rules~\cite{das94} express the fact that seemingly different operators 
are not necessarily linearly independent on a given SU(6) representation. 
For example, $\widehat{\cal O}_3$ and ${\cal O}_2$ are linearly dependent 
when applied to the spin-flavor symmetric ground state ${\bf 56}$-plet.

The existence of unique one-, two-, and three-quark operators can also 
be understood from the following group theoretical argument. 
From the viewpoint of broken 
SU(6) spin-flavor symmetry both the $N$ and $\Delta$ belong to the same 
${\bf 56}$ dimensional ground state 
multiplet. Therefore, an allowed symmetry breaking operator $\cal{O}$
must transform according to one of the irreducible representations 
found in the product
\be
\label{rep}
\bar{{\bf 56}} \times {\bf 56}
=  {\bf 1} + {\bf 35} + {\bf 405} + {\bf 2695}.
\ee
Here, the ${\bf 1}$, ${\bf 35}$, ${\bf 405}$, and
${\bf 2695}$ dimensional representations,
are respectively connected with zero- (a constant), one-, two-,
and three-body operators. Because each representation 
on the right-hand side of Eq.(\ref{rep}) occurs only once,
the operators in Eq.(\ref{operators}) are unique in the sense that 
for each ${\cal O}_i$ there is only one linearly independent operator 
structure. As a result, the operators in Eq.(\ref{operators}) provide a 
complete spin-flavor basis for the observables considered here.

\begin{table}[htb]
\begin{tabular}{|l|c|c|c|} \hline
baryon & ${\cal M}_1$ & ${\cal M}_2$ & ${\cal M}_3$ \\ \hline
p & $\frac{5}{3}A_1$ & -$\frac{2}{3} A_2$ &  -$\frac{26}{3} A_3$ \\
$p \rightarrow \Delta^+ $ & $ \frac {4 \sqrt{2}}{3} A_1$ & 
- $ \frac {4 \sqrt{2}}{3}A_2$ & $ \frac {8 \sqrt{2}}{3}A_3$  \\ 
$\Delta^{+}$ & $    A_1$ & $ 2 A_2$ & $2 A_3$ \\ 
\hline
\end{tabular}
\caption[Table 1]{\label{tab1} Matrix elements ${\cal M}_{i}$ 
of the one-quark (${\mathcal O}_1$), two-quark (${\mathcal O}_2$), 
and three-quark (${\mathcal O}_3$) 
axial vector operators in Eq.(\ref{operators}). 
The matrix element of the total operator ${\cal O}$ in Eq.(\ref{quarkop}) 
is ${\cal M}_q={\cal M}_{1}+{\cal M}_{2}+{\cal M}_{3}.$}
\end{table}

\section{Results}
Evaluating the operators in Eq.(\ref{operators}) between SU(6) 
wave functions~\cite{Clo} 
for the $N$ and $\Delta$  we get the results 
compiled in Table~\ref{tab1}.

To obtain from the total quark level matrix elements ${\cal M}_q$ 
in Table~\ref{tab1} the conventional pion-baryon couplings, 
one proceeds as follows~\cite{Bro75}. 
Conventionally, the $N\to \Delta$ transition vertices  
depicted in Fig.~\ref{fig:deltacoup} (left) are defined as matrix elements 
${\cal M}_B$  of baryon level $N \to \Delta$ transition 
spin ${\bf S}$ and isospin 
${\bf T}$ operators. The latter are normalized so that their matrix elements 
are equal to the corresponding spin and isospin Clebsch-Gordan coefficients  
\bea 
\label{conv1}
{\cal M}_B & = & f_{\pi N\Delta} \, 
\langle  \Delta, S'\, S'_z \, T'\, T'_z \vert {\bf S}_z \, {\bf T}_z 
\vert N, S\, S_z,\, T\, T_z \rangle \nonumber \\
& = &  f_{\pi N \Delta} \, 
(1\, 0 \  S \, S_z \vert S' \,S_z')\,\, (1\, 0 \  T\, T_z \vert T'\, T_z').
\eea
Here, $S=T=1/2$ refers to the spin and isospin of the $N$, and 
$S'=T'=3/2$ to the spin and isospin of the $\Delta$.
For the $p\to \Delta^+$ transition with
$T_z=T'_z=1/2$ and $S_z=S'_z=1/2$ Eq.(\ref{conv1}) gives  
${\cal M}_B = f_{\pi^0 p \Delta^+} (2/3)$.
This baryon level matrix element must be equal to the 
corresponding total quark level matrix element, i.e.,  
${\cal M}_B=f_{\pi^0 p \Delta^+} (2/3)={\cal M}_q$. 
Consequently, the $f_{\pi^0 p \Delta^+}$ coupling is obtained 
by multiplying the total quark level matrix element  ${\cal M}_q$ 
in the second row of Table~\ref{tab1} by 3/2.
 
Analogously, $\Delta \Delta$ vertices
depicted in Fig.~\ref{fig:deltacoup} (right) are obtained from the 
diagonal matrix elements
\bea 
\label{conv2}
{\cal M}_B & = & f_{\pi \Delta \Delta} \, 
\langle  \Delta, S'\, S'_z\,  T'\, T'_z \vert {\bf S}_z \, {\bf T}_z 
\vert \Delta, S'\, S'_z \, T'\, T'_z \rangle \nonumber \\
& = &  f_{\pi \Delta  \Delta} S'_z \,  T'_z. 
\eea
Evaluating Eq.(\ref{conv2}) for the $\Delta^+$ with $T'_z=1/2$ 
and maximal spin projection $S'_z=3/2$
gives ${\cal M}_B = f_{\pi^0 \Delta^+ \Delta^+} (3/4)$.
This baryon level matrix element must be equal to the corresponding 
total quark level matrix element, i.e., 
${\cal M}_B= f_{\pi^0 \Delta^+ \Delta^+} (3/4) = {\cal M}_q$.  
Therefore, the entry for $\Delta^{+}$ in Table~\ref{tab1} 
must be multiplied by  $4/3$.

We then get 
\bea
\label{result1}
f_{\pi^0 p p} & = & \frac{5}{3}\, A_1 - \frac{2}{3}\, A_2 - 
\frac{26}{3}\, A_3, \nonumber \\
f_{\pi^0 p \Delta^+} & = & \frac{4\sqrt{2}}{3}\, 
\left ( A_1 - A_2 + 2 A_3 \right ) \, \left (\frac{3}{2} \right ),
\nonumber \\
f_{\pi^0 \Delta^+ \Delta^+} & = & 
\left (A_1 +  2 \, A_2 + 2 \, A_3 \right )
\left (\frac{4}{3} \right ).
\eea
Solving Eq.(\ref{result1}) for the constants $A_i$ leads to 
\bea
\label{result2}
A_1 & = & \frac{1}{6} \, f_{\pi^0 p p} 
+ \frac{\sqrt{2}}{9} \, f_{\pi^0 p \Delta^+} 
+ \frac{5}{24}\, f_{\pi^0 \Delta^+ \Delta^+}, \nonumber \\
A_2 & = & \frac{1}{4}\, f_{\pi^0 \Delta^+ \Delta^+} 
- \frac{\sqrt{2}}{12}\, f_{\pi^0 p \Delta^+}, \nonumber \\
A_3 & = & -\frac{1}{12} \, f_{\pi^0 p p} 
+ \frac{\sqrt{2}}{36} \, f_{\pi^0 p \Delta^+} 
+ \frac{1}{48}\, f_{\pi^0 \Delta^+ \Delta^+}. 
\eea

Next, we calculate numerical values for the strong $\Delta$ couplings,
including successively first, second, and third 
order SU(6) symmetry breaking terms represented respectively
by the operators ${\cal O}_1$, ${\cal O}_2$, and 
${\cal O}_3$ in Eq.(\ref{operators}).

First, ignoring two- and three-quark terms, 
we get $A_1 =(3/5)\, f$, 
where we use the abbreviation $f:=f_{\pi^0 p p}$.
Thus, $A_1$ is fixed by the empirical value for the 
strong $\pi NN$ coupling ${f_{\pi^0 p p}^2}/({4 \pi}) =0.08$. 
In this first order SU(6) symmetry breaking approximation, we reproduce 
the well known additive quark model results 
for the $\pi \Delta$ couplings~\cite{Bro75}  
\bea
\label{impulse}
f_{\pi N\Delta} & = & \frac{6\sqrt{2}}{5}\, f,  \nonumber \\
f_{\pi \Delta \Delta} & = & \frac{4}{5}\, f.
\eea

Second, if we include two-quark but still neglect three-quark terms 
we need the empirical relation $f_{\pi N \Delta}=2.1 \, f$ to fix 
the additional constant $A_2$. 
In this case, SU(6) symmetry is broken up to second order.
Eq.(\ref{result1}) with $A_3=0$ gives then
$A_1=0.51 \, f$ and $A_2= -0.24 \,f$
as in Ref.~\cite{hen00}. 
In this approximation we recover Eq.(\ref{rel1}), which as shown 
in Table~\ref{tab2} entails an unrealistically small value 
for $f_{\pi^0 \Delta^+ \Delta^+}$.

Finally, the inclusion of three-quark terms takes
third order SU(6) symmetry breaking into account.
Using the QCD sum rule value 
$f_{\pi^0\Delta^+\Delta^+}=0.666 \, f$~\cite{zhu00},
which is consistent with the data~\cite{arn79,Kam04,sko11}, 
allows us to fix the constant $A_3$ in Eq.(\ref{result1}). We 
then get from Eq.(\ref{result2}) the follwing values for the 
constants: $A_1=0.635 \, f$, $A_2= -0.081 \, f$, and $A_3=0.013\, f$.
By taking three-quark operators into account, 
we find that Eq.(\ref{rel1}) is modified as follows
\be
\label{rel2} 
f_{\pi^0 p p}- \frac{1}{4}\,  f_{\pi^0 \Delta^+\Delta^+} =
\frac{\sqrt{2}}{3} \, f_{\pi^0 p \Delta^+} - 12 A_3.
\ee
Consequently, $f_{\pi^0 \Delta^+\Delta^+}$ can be of the same magnitude as 
$f_{\pi^0 pp }$ even when the empirical
value for $f_{\pi^0 p \Delta^+}$ is used so that the discrepancy
between theory and experiment can be resolved. 

\begin{table}[htb]
\begin{tabular}{|l|c|c|c|c|c|} \hline
coupling &${\cal M}_1$ & ${\cal M}_1 + {\cal M}_2$  & 
${\cal M}_1 + {\cal M}_2 + {\cal M}_3 $ & exp.  \\ 
\hline
$f_{\pi^0 pp}$ & 1.00 & 1.00 & 1.00 & 1.00~\cite{eri88}\\ 
$f_{\pi^0 p \Delta^+}$ & 1.70  & 2.10 & 2.10 & 2.10~\cite{eri88} \\
$f_{\pi^0 \Delta^+ \Delta^{+}}$  & 0.80 & 0.04 & 0.67 
& 0.06-1.72~\cite{arn79}\\
\hline
\end{tabular}
\caption[Table 2]{\label{tab2}
Pion couplings of the $\Delta(1232)$ in terms of 
the pion-nucleon coupling $f=f_{\pi^0 p p}$ with the successive
inclusion of one-quark (${\cal M}_1$), 
two-quark (${\cal M}_2$),  
and three-quark (${\cal M}_3$) terms 
compared with experimental data.
The experimental range for $f_{\pi^0 \Delta^+\Delta^+}$ is from Table 9 
in Ref.~\cite{arn79}.}
\end{table}

Table~\ref{tab2} lists the strong couplings
$f_{\pi^0 p p}$, $f_{\pi^0 p \Delta^+}$, and $f_{\pi^0 \Delta^+ \Delta^+}$ 
in terms of $\,f\,$ in the successive approximations discussed above.
The numbers in the first column  
correspond to the additive quark model results in Eq.(\ref{impulse}). 
The entries in the second column include 
the effect of the two-quark operator ${\cal O}_2$ of Ref.~\cite{hen00}.
The latter changes the double $\Delta$ coupling from the additive quark model 
value $f_{\pi\Delta\Delta}=(4/5)f$
to $f_{\pi\Delta\Delta}=0.04 f$. 
However, such a small value for $f_{\pi\Delta\Delta}$ is 
inconsistent with other QCD based baryon structure models,  
which predict that $f_{\pi\Delta\Delta}$, $f_{\pi N\Delta}$, and $f_{\pi NN}$ 
are of the same order of magnitude.  
Finally, the third column  
represents a full calculation of one-, two-, and three-quark contributions. 
With three-quark terms included, the 
double $\Delta$ coupling changes from $0.04 f$ to $0.67 f$,
in qualitative agreement with QCD sum rule~\cite{zhu00} and $1/N_c$ 
expansion~\cite{das94} calculations. This provides 
evidence for the importance of three-quark operators in 
axial vector quantities, such as the pion-baryon couplings.

It is of interest to compare our results to those of Abbas~\cite{aba91}, 
who found the following expressions for the $\pi N \Delta$ and 
$\pi \Delta \Delta$ couplings based on one-quark axial vector operators 
but with $D$ state admixtures in the $N$ and $\Delta$ wave functions 
\bea
\label{abbas}
f_{\pi N\Delta} & = & \frac{6\sqrt{2}}{5}\,
\left \lbrack\frac{1-\frac{1}{2}\,P_D}{1-\frac{6}{5}\, P_D} \right \rbrack  
\, \frac{1}{\sqrt{1+P_D}} \, f,  \nonumber \\
f_{\pi \Delta \Delta} & = & \frac{4}{5}\, \left \lbrack 
\frac{1- P_D}{1-\frac{6}{5}\, P_D} \,\, \frac{1}{1+P_D} \right \rbrack f,  
\eea
where $P_D$ is the $D$-state probability in the nucleon 
wave function~\cite{comment2}.
Note that the double $\Delta$ coupling in Eq.(\ref{abbas}) has been 
adjusted to the normalization convention used here~\cite{comment1}. 
For $P_D=0.34$ one finds
$f_{\pi N \Delta}=2.1 f$ 
and  $f_{\pi \Delta \Delta}=0.67 f$.
Incidentally, the same value $P_D=0.34$ 
also explains the empirically small quark contribution to 
nucleon spin~\cite{aba89}. 

As a result of the large $D$-state admixture in the nucleon 
and $\Delta$ wave functions, $f_{\pi N \Delta}$ is increased by 20$\%$,   
and $f_{\pi \Delta \Delta}$ is decreased by the same percentage  
with respect to the additive quark model values in Eq.(\ref{impulse}).
This is consistent with the data and in quantitative agreement 
with the present results. Apparently, nonspherical $N$ and $\Delta$ 
wave functions with large $D$-state probabilities have the same effect as the 
two- and three-quark contributions considered here.

This correspondence between large $D$ wave admixtures and
many-quark operators is not a coincidence. 
Via a unitary transformation it is always possible to eliminate 
many-body operators at the expense of more complicated wave functions
without changing the observable matrix element~\cite{fri75}. 
Yet, the inclusion of many-quark operators leads overall 
to a more consistent and realistic description of nucleon structure
for the following reasons. First, ab initio quark model 
calculations based on gluon and pion exchange potentials feature much 
smaller $D$ state probabilities ($< 0.01$)~\cite{buc91} for the $N$ and
$\Delta$ wave functions. Second, many-quark operators are related 
to the quark-antiquark degrees of freedom, commonly refered to as meson cloud 
in physical baryons. Baryon deformation is more likely  a result of 
these non-valence quark degrees of freedom than a consequence 
of massive valence quarks moving in elliptical orbits~\cite{hen01}. 

In any case, the ability of both models to describe 
the strong $\pi \Delta$ couplings is closely tied 
to nonspherical geometric shapes of the nucleon and $\Delta$. 
Other observables, such as the $N \to \Delta$ quadrupole transition 
moment, baryon octupole moments, and the quark contribution to 
nucleon spin~\cite{hen01} point to the same conclusion concerning 
the nonsphericity of both $N$ and $\Delta$ states.

\vspace{1 cm}

\section{Summary}
In summary, we found that previously uncalculated three-quark operators 
make a significant contribution to the $\pi N \Delta$ 
and $\pi \Delta \Delta$ coupling constants.
In particular, for the double $\Delta$ coupling, the  
three-quark term reduces the influence of the negative two-quark 
contribution so that the final result $f_{\pi^0 \Delta^+\Delta^+}=0.67\, f$ 
is about $20\%$ smaller than the additive quark model value
consistent with data and other QCD based baryon structure models.

Furthermore, our theory is in quantitative  agreement 
with results obtained in a quark model with a large $D$-state admixtures. 
Both approaches increase $f_{\pi N \Delta}$ 
and simultaneously decrease $f_{\pi \Delta \Delta}$
by about $20 \%$ with respect to the additive quark 
model. This indicates that nonspherical $N$ and $\Delta$ states are
necessary for a quantitative understanding of the strong pion couplings
independent of whether the spatial deformation is described as 
large $D$ state components in the valence quark wave functions
or (more realistically) as two- and three-quark operators 
representing a nonspherical sea of quark-antiquark pairs.  

Having demonstrated the importance of two- and three-quark terms 
for a consistent description of the $\pi N$ and $\pi \Delta$ couplings 
it will be interesting to investigate their effect on other 
axial vector quantities, e.g. the $p \to \Delta^+$ 
transition magnetic moment and the weak axial $N \to \Delta$ 
transition~\cite{zhu02} for which large discrepancies
between the additive quark model and experiment persist.


\begin{thebibliography}{99}
\bibitem{comment1} 
Other authors, e.g. Refs.~\cite{gre76,aba91}, use the normalization 
convention $f_{\pi \Delta \Delta} =(1/5) \, f_{\pi NN}$.
This is obtained if 
$f_{\pi^0\Delta^+\Delta^+}$ is evaluated for $\Delta^+$ spin projection 
$S'_z=1/2$, which gives
$\langle \Delta^+, S'_z\!\!=\!\!1/2 \,
\vert \, {\cal O}_1 \, \vert \, \Delta^+, S'_z\!\!=\!\!1/2 \rangle = A_1/3$, 
instead of 
$\langle \Delta^+, S'_z\!=\!3/2 \,
\vert \, {\cal O}_1 \, \vert \, \Delta^+, S'_z\!=\!3/2 \rangle = A_1$ 
for maximal spin projection $S'_z=3/2$ 
as in Table~\ref{tab1}. 
In our normalization scheme  (see sect.3) 
the $\pi \Delta$ couplings are independent of the 
considered spin and isospin channel.
\bibitem{comment2} 
The $D$ state probability in the $\Delta$ 
wave function $P_D(\Delta)$ is related to the $D$ state probability 
in the nucleon wave function 
$P_D$ as $P_D(\Delta)= 2 P_D/(1+P_D)$~\cite{aba91}.
This has been used to eliminate $P_D(\Delta)$ from the 
original expressions for the $\pi \Delta$ couplings.
\end{thebibliography}
\end {document}